\documentclass[preprint, 12pt]{aastex}
\usepackage{epsfig, amsmath, amsfonts}

\newcommand\sol{ {\odot }}

\newcommand{\dd}[2]{\frac{{\rm d} #1}{{\rm d} #2}}

\def\dd{\partial}

\def\beq{ \begin{equation} }
\def\eeq{ \end{equation} }
\def\spose#1{\hbox to 0pt{#1\hss}}  
\def\ltsim{\mathrel{\spose{\lower.5ex\hbox{$\mathchar"218$}}
\raise.4ex\hbox{$\mathchar"13C$}}}
\def\gtsim{\mathrel{\spose{\lower.5ex\hbox{$\mathchar"218$}}
\raise.4ex\hbox{$>$}}}

\slugcomment{}

\begin{document}

\title{\bf\LARGE The Effect of the Tachocline on Differential Rotation in the Sun}
\author{ Steven A. Balbus\altaffilmark{1,2}, Henrik N. Latter\altaffilmark{1}}
\altaffiltext{1}{Laboratoire de Radioastronomie, \'Ecole Normale
Sup\'erieure, 24 rue Lhomond, 75231 Paris CEDEX 05, France
  \texttt{steven.balbus@lra.ens.fr}}

  \altaffiltext{2}{Adjunct Professor, Department of Astronomy, University of Virginia,
  Charlottesville VA 22903, USA}

\begin{abstract}
In this paper, we present a model for the 
effects of the tachocline on the differential rotation in the
solar convection zone.  The mathematical technique relies
on the assumption that entropy is nearly constant (``well-mixed'')
in isorotation surfaces both outside and within the tachocline. 
The resulting solutions exhibit nontrivial features that strikingly resemble
the true tachocline isorotation contours in unexpected detail.   
This strengthens the mathematical premises of the theory.  
The observed rotation pattern in the tachocline shows 
strong quadupolar structure, an important feature that is explicitly used
in constructing our solutions.  
The tachocline is treated locally as an interior boundary layer of small but
finite thickness, and an explicit global solution
is then constructed.  A dynamical link can thus be established between
the internal jump in the angular velocity at the tachocline and
the spread of angular velocities observed near the solar surface. 
In general, our results suggest that the bulk of the solar convection zone is
in thermal wind balance, and that simple quadrupolar stresses,
local in radius, 
mediate the tachocline transition from differential rotation to
uniform rotation in the radiative interior.

\end{abstract}

\keywords{convection --- hydrodynamics ---
stars: rotation --- Sun: rotation --- Sun:
helioseismology}

\maketitle
\section{Introduction}

Recent work suggests that the isorotation contours in the bulk of
the solar convection zone (hereafter SCZ), 
correspond to the characteristic
curves of the vorticity equation in its ``thermal wind'' form
(Balbus 2009, hereafter B09;
Balbus et al. 2009, hereafter BBLW; Balbus \& Weiss 2010).  
In its undeveloped form,
this equation relates entropy gradients to angular velocity gradients.
The mathematical existence of
the characteristics demands, therefore, 
some sort of functional relationship between the entropy $S$ and 
angular velocity $\Omega$.  
The theory presented in the above papers is predicated upon the notion
that within constant $\Omega$ surfaces, the dominant, long-lived convective
cells carry out their task of mixing entropy with great efficiency.
This means that within a constant $\Omega$ surface, the excess entropy gradient
above and beyond the minimal radial gradient 
needed to maintain the convective state is, in essence,
eliminated.   The remaining ``residual'' entropy is thus 
uniform in constant $\Omega$ surfaces.  More explicitly, the
residual entropy
$S'\equiv S-S_r$ is formed by subtracting a function of spherical
radius $r$ (something close to the driving background entropy
that has been averaged over angles), and $S'$ is set equal to a
function of $\Omega$.  This is a 
powerful constraint, providing the necessary mathematical link for
interpreting the solar isorotation contours as trajectory
characteristics of a quasilinear partial differential equation.  
BBLW advances several different lines of argument (heuristic, numeric,
and goodness-of-fit) supportive of this
picture.  Originally
an MHD basis for the $S-\Omega$ connection was 
put forth (B09), but the greater simplicity and generality, as well
as the broad
agreement with many different lines of argument, together suggest
primacy for the BBLW mechanism.  

An attractive feature of the theory presented in BBLW is that it provides simple
and intuitive answers to not-so-trivial questions.  For example, why can't
the solar convection zone rotate on cylinders?  Because if
it did, then $\Omega$ would necessarily depend upon the spherical 
colatitude angle
$\theta$, therefore the entropy would also depend upon $\theta$, and this
is inconsistent with a vanishing $z$ gradient in the thermal
wind equation (TWE; cf. equation [2] below).
Why then is $\Omega$ dominated
by its $\theta$ gradient?   Now we argue the other way: because the residual 
entropy has its dominant radial gradient removed (physically by
mixing, mathematically by a simple subtraction), 
the relative
importance of its $\theta$ gradient has been enhanced.  But residual entropy shares
the same isosurfaces as $\Omega$.  In numerical simulations, the fact that
imposed latitudinal entropy gradients enforced at an interior
boundary lead also to latitudinal
gradients in $\Omega$ (Miesch, Brun, \& Toomre 2006)
becomes tautological: latitudinal $S$-gradients lead inevitably to
latitudinal $\Omega$-gradients because these gradients
are always counteraligned.  

In its current formulation,
BBLW theory addresses the differential rotation in the Sun away
from the tachocline.  This is nothing if not prudent:           
the dynamics of the solar tachocline is a 
notoriously controversial and difficult topic
(e.g. Hughes, Rosner, \& Weiss 2007). 
Yet, in studying the isorotation contours within the tachocline, one is
struck both by the highly localized nature of the distortion, and by the
relative simplicity and uniformity of the their appearance.  In contrast
to the bulk of the convection zone and the outer boundary layer, there is a sense
of inevitability to these curves.  
Perhaps their gross form might
not be sensitive to the complex details of the 
internal dynamical structure of the
tachocline, any more than the gross adiabatic temperature profile
of the SCZ depends upon the details of the convective heat
flux.  We might look instead to the exigencies of local forcing
and to the strikingly quadrupolar structure of the region.
The resulting mathematical demands of joining the isorotation contours
in such a forced region to 
the overlying convective zone contours treated in BBLW may
be sufficiently restrictive
that the role of turbulence is tightly confined.  In its current form,
BBLW theory has too much ``free play:''  it needs to be restrained.  
In this work, we will develop this point of view, and pursue its 
consequences.

\section{Analysis}
\subsection{Trajectory and solution characteristics of the TWE}

We follow the notation convention of BBLW.  
Cylindrical coordinates are denoted by radius $R$, azimuthal angle
$\phi$ and vertical coordinate $z$.  The spherical coordinates 
are $(r, \theta, \phi)$, where $r$ is the radius
from the origin, $\theta$ the colatitude angle, and $\phi$ 
the azimuthal angle.   The angular velocity $\Omega$,
pressure $P$ and density $\rho$ 
are understood to be azimuthal averages.
The dimensionless entropy function $\sigma$ is defined by:
\beq
\sigma \equiv \ln P\rho^{-\gamma},
\eeq
where $\gamma$ is the usual specific heat ratio, or adiabatic index.  

In SCZ applications, the TWE is the dominant balance of the vorticity
equation, after azimuthal averaging, 
between the large scale rotation and
the baroclinic terms.  The latter arises from a 
mismatch between equipotential
(or isobaric) and isochoric surfaces.  
More specifically, contributions from convective turbulence are
ignored (i.e., the
convective Rossby number is small [Miesch \& Toomre 2009])
as are those from from magnetic fields.  
The TWE for the convective zone is (e.g. Thompson et al. 2003, 
Miesch 2005, B09):
\beq\label{eq1}
R{\dd\Omega^2\over \dd z}={g\over\gamma r}
{\dd \sigma
\over \dd\theta},
\eeq
where $g=GM_\odot/r^2$, $G$ is the gravitational
constant and $M_\odot$ is a solar mass.  In BBLW theory, 
there is a functional relationship of the form
\beq\label{eq1bis}
\sigma' \equiv \sigma -\sigma_r =f(\Omega^2)
\eeq
where $\sigma_r$ is any function of $r$ (in practice something very close to an
angle-averaged $\sigma$) and $f$ is an unspecified function.   
Replacing $\sigma'$ by $\sigma$ does not alter the TWE, and
with the aid of equation (\ref{eq1bis}), the TWE (\ref{eq1}) 
becomes
\beq\label{eq2}
{\dd\Omega^2\over \dd r} - 
\left(
{gf'\over\gamma r^2\ \sin\theta \cos\theta}   +
{\tan\theta\over r}
\right)
{\dd\Omega^2\over \dd\theta} = 0 ,
\eeq
where $f'=df/d\Omega^2$.  In this study, we wish to retain,
at least in a formal sense, the tachocline meridional stress $T(r,\theta)$ on the
right side of the equation    
\footnote{
In passing from the lower convective zone into the radiative layers,
the baroclinic term proportional to $(\dd P/\dd\theta)(\dd \sigma/\dd r)$
should be formally be retained in our equation.  As a practical matter,
the rotational pattern is essentially spherical in this region, and the 
dominant balance is between the inertial terms and the external stress.
(We thank the referee M. Miesch for drawing our attention to this point.)}:
\beq
\label{eq3}
{\dd\Omega^2\over \dd r} -
\left(
{gf'\over\gamma r^2\ \sin\theta \cos\theta} +
{\tan\theta\over r}
\right)
{\dd\Omega^2\over \dd\theta} = T(r,\theta).
\eeq
Although $T$ is certainly poorly understood in detail, there is 
presumably {\em some} function of position that makes
equation (\ref{eq3}) correct.  The point is that
the ensuing formal solution is not without content.   The angular velocity
$\Omega^2$ satisfies
\beq\label{sol1}
{d\Omega^2\over dr} = T(r,\theta)
\eeq
on 
\beq\label{tra1}
{d\theta\over dr } = - 
{gf'(\Omega^2)\over\gamma r^2\ \sin\theta \cos\theta}    -
{\tan\theta\over r}.
\eeq
In other words, to solve for $\Omega(r,\theta)$, we use 
the contours as defined by the characteristic
differential equation of BBLW theory; along these contours
$\Omega$ changes according to equation (\ref{sol1}).

We shall assume that equation (\ref{eq1bis}) is valid both in the 
bulk of the convective zone as well as in the tachocline.  This 
is by no means obvious.  The low Rossby number approximation is
likely to hold well in the tachocline and convective rolls
should be strongly affected by shear.  But since the turbulent stresses
in the tachocline are very different from those in the convective zone,
even if such a functional relationship holds, the function $f$ may in 
principle be quite different in the two regions.  

On the other hand {\em everything} changes in the two regions, not the
least of which is the sudden
appearance of the stresses themselves.   For the class of solutions we consider here
(which reproduce the data strikingly well), a change in $f'$ cannot be disentangled from
a change in $T$, a point we will explicitly discuss (\S 2.6).  Hence, we retain
the functional equation (\ref{eq1bis}) throughout the convective zone and the 
tachocline.  

Note that, in contrast to the vanishing $T$ limit, when $T$ is finite
$f'(\Omega^2)$ is generally {\em not} constant along the characteristic.
But this does not mean that there is nothing to be learned from considering
the constant $\Omega$ as a special case for the current, more general, problem.  
Indeed, in our first approach we should keep formalism to a minimum.  
It has been earlier noted (B09)
that choosing $f'(\Omega)$ to be a global constant captures
important semi-quantitative features of the true SCZ isorotation
contours, so let us begin first with this mathematically simple case.
We will then be poised to look at a more complex problem.  

\subsection {$f'(\Omega^2) =$ constant}

In the limit of constant $f'$, the formal
trajectory characteristics are identical to those
in BBLW theory, both outside and within the tachocline:
\beq\label{tra2}
\sin^2\theta = {r_0^2\over r^2} \sin^2\theta_0 + {2f'GM_\odot\over r^2
\gamma}
\left( {1\over r} - {1\over r_0}\right),
\eeq
where $r_0$ is the radius at which $\Omega$ is initially specified
(generally near or at the surface),
and $\theta_0$ marks the beginning value of $\theta$ for a
particular characteristic path.  
The sign of $T$ is 
now clear: since along a BBLW characteristic, $\Omega$ increases with decreasing 
$r$, $T$ must be negative at high latitudes and positive at equatorial latitudes.
Along such a characteristic, denoted $\theta(r)$,  $\Omega^2$ is given by
\beq\label{sol2}
\Omega^2 = \Omega_0^2+ \int_{r_0}^{r} T\big[r, \theta(r)\big]\, dr
\eeq
where $\Omega_0$ is the value of $\Omega$ at the defining surface.
(Note that $r_0$, which is generally taken at or near the
surface, may be larger than $r$.)   The formal solution (\ref{sol2}) is
in fact general for any $f'$, provided that $\theta(r)$ is taken to be
the proper trajectory characteristic.  

It is not difficult to extract the qualitative behavior of 
$\Omega$ from
equations (\ref{tra2}) and (\ref{sol2}).  The stress 
$T$ is very small except in a very narrow region near the
transition radius $r_T$.  Thus $\Omega^2$ remains essentially
fixed 
at its initial value $\Omega_0^2$ along most of the extent of the characteristic.
This, of course, is BBLW theory.  Then, there is a sudden increase in $T$ very
near $r_T$, and along the BBLW characteristic $\theta(r)$,
$\Omega$ now makes a very sharp rise (or a sharp drop near the equator)
as it settles to the
uniform rotation rate in the radiative interior. 

The resulting isorotational curves can be understood with the 
aid of figure (1).  This diagram is a local representation
of the SCZ-tachocline boundary in a region where
the angular velocity begins to increase sharply
moving inward.   The radius $r$ increases upward to the right;
$\theta$ increases downward to the right. 
The numbers $1'$ though $4'$
label BBLW characteristics,
which in the convection zone are precisely the isorotation contours.
Hence these numbers may also be thought of as angular velocities,
increasing with the magnitude of the number.  

Consider the rightmost contour labeled $4'$.
Between $4'$ and the dot labeled 4 on the same curve, the contour passes
through the
convective zone, and the angular velocity remains constant.
Dot 4 marks the beginning of the tachocline, and a steep rise
in the angular velocity.  To continue the trace of the isorotation
contour, a sharp turn to the (reader's) left is required,
and the curve continues by passing through dot 4 on the curves
labeled $3'$, $2'$, and $1'$, etc.  Thus,
within the narrow radial band of changing $\Omega$, 
if one were to follow curves of constant $\Omega$, from 
one BBLW trajectory characteristic to those immediately adjacent, 
a quasi-spherical ($r$ increases somewhat with $\theta$)
path would emerge.  This abrupt change in the isorotation contours
is clearly seen in the helioseismology
data.   The more delicate $\theta$ dependence of $r$ is also present
in the data, though this may be an artifact
of the inversion technique.  

\begin{figure}
\centerline{\epsfig{file=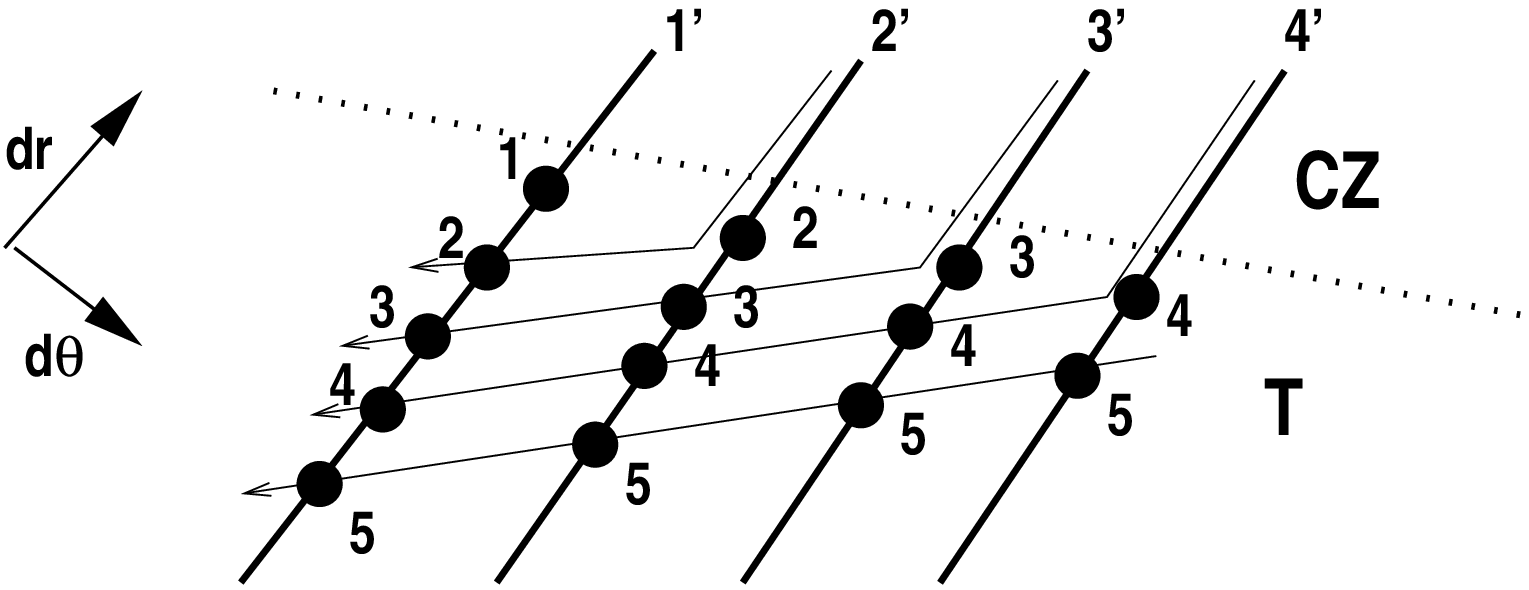, width=12 cm}}
\caption{Schematic local
rendering of the isorotation contours in the presence of
tachocline meridional stress.  CZ denotes the convective
zone and T denotes the tachocline.  
The thick curves labeled $1', 2',$ etc.\ are
the CZ isorotation contours in BBLW theory.  Numbers
may be thought of as representing the relative ordering of
fiducial $\Omega$ values.  The presence of the tachocline,
indicated by black dots, is
to increase $\Omega$ along what was an isorotation
contour.  The new, proper isorotation contours
(thin lines) resemble those seen in the helioseismolgy data: 
quasi-spherical, 
with an increase in $r$ as $\theta$ increases, and with a         
sharply angled upward turn toward the surface.}
\end{figure}

\subsection {Green's Function Solution}

The above description motivates a Green's function approach.
Consider a tachocline stress of the form
\beq
T(r,\theta) = - (\Delta\Omega^2)F(\theta)\delta(r - r_T)
\eeq
where $\Delta\Omega^2$ is the ``jump amplitude,'' $F(\theta)$ is
a function of $\theta$ that modulates the magnitude of the radial inward jump
($F$ is positive at high latitudes, negative at small latitudes),
and $\delta(r-r_T)$ is the Dirac delta function.  Then along
a BBLW characteristic,
\beq\label{omg}
\Omega^2 = \Omega_0^2 +
\Theta(r_T - r)
(\Delta\Omega^2)F\big(\theta(r_T)\big)
\eeq
where $\theta(r)$ may be determined explicitly from equation (\ref{tra2}),
and 
$\Theta(x)$ is the Heaviside function.  (Recall
that $\Theta$ is unity when its argument is positive and zero
when its argument is negative.)  In this rather drastic
idealization,
BBLW theory breaks down at a single radius $r_T$, which we identify
with the tachocline.  The BBLW isorotation contours
change abruptly to a single spherical shell in this model,
compressing the entire tachocline down to a range of vanishing thickness.

\subsection {Quadrupole Structure}

It is possible, of course, to continue along the lines of
the previous section by introducing
progressively more sophisticated models for
$T(r,\theta)$ and $f'(\Omega^2)$, and in fact we shall shortly consider
the case in which $f'$ is a linear function of $\Omega^2$.  
It is already clear, however, that within this scheme,
the helioseismology data
can be reproduced---at least qualitatively. 
If our approach is not quite tight enough 
to be broadly predictive, it might be more
profitably applied in the reverse sense:  use the solar rotation
data directly to infer the form of $T(r,\theta)$.

\begin{figure}
\begin{center}
\epsfig{file=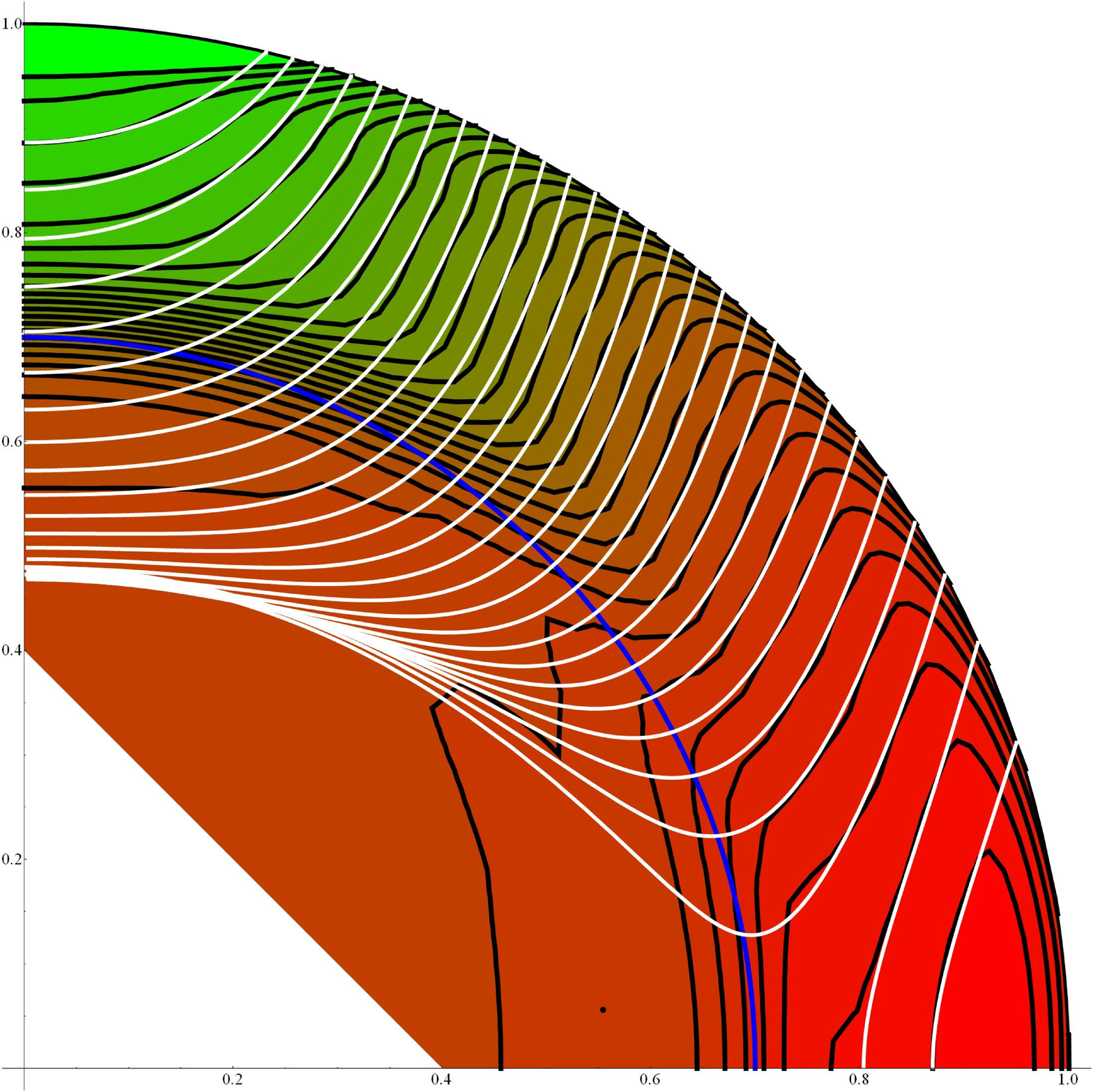, width=8cm}
\epsfig{file=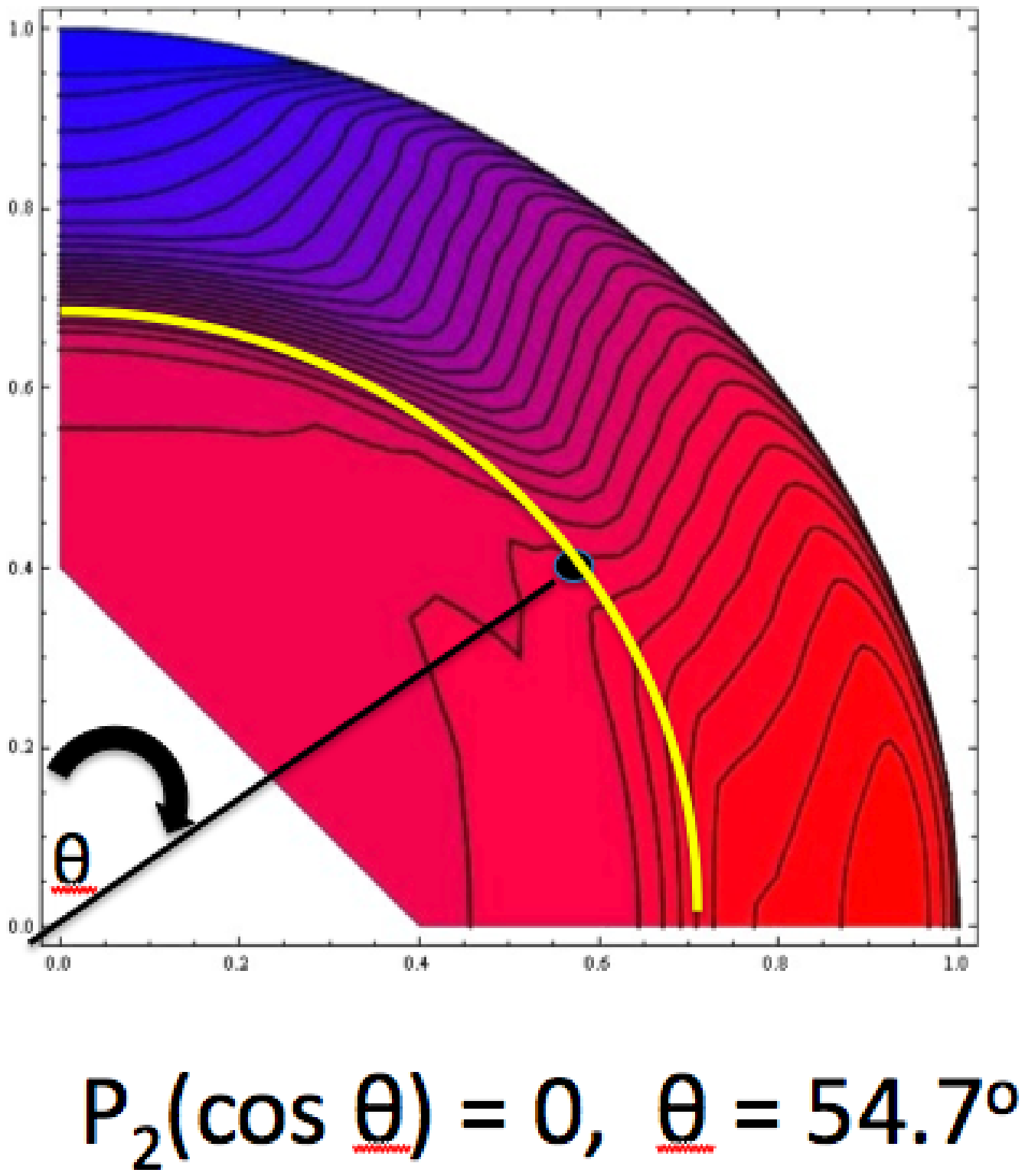, width=8 cm}
\end{center}
\caption {Left: Constant $\Omega$ contours 
(white) of the thermal wind equation from BBLW theory
overlaid on top of (black) isorotation contours 
(GONG data, courtesy of R. Howe).  Blue curve
is the bottom edge of the convective zone.  Right: Tachocline is
absent when $P_2(\cos\theta)=0$, consistent with quadrupole structure.
}
\end{figure}

A very important result emerges immediately.  In figure (2),
on the left we see 
the isorotation contours taken from GONG data\footnote{We thank
R. Howe for providing us with these results.}, together 
with the BBLW characteristics overlaid for reference.
As one passes through the tachocline from larger to smaller radii,
the inward jump in $\Omega^2$ changes 
from positive to negative for 
$\theta$ larger than $54.7^\circ$, the value of $\theta$ for
which $\cos^2\theta=1/3$.  The inset to the right in figure (2),
displaying GONG data together
with the location of the critical angle, shows this
very clearly.  The magnitude of the jump
then grows (in the negative sense) as $\theta$ 
increases, as we move toward the equator.
The spacing of the isorotational contours is, at least crudely,
about half the interval at the poles as compared with the equator.
Even granting the inevitable uncertainties associated with 
the polar data (which are not reliable poleward of $70$ degrees),
these results are much what one would 
expect if $F(\theta)$, the angular component in our radially local
model, were proportional to 
$$P_2(\cos\theta) = {1\over2}(3\cos^2\theta - 1),$$
the Legendre
polynomial of order 2.   To leading order, {\em the tachocline stresses
are revealing a dominant quadrupolar structure.} 

This is such a fundamental property of tachocline structure, it is curious
that the near coincidence of the zero of $P_2(\cos\theta)$
and the ``zero'' of the tachocline has not been emphasized in the earlier literature.
While meridional flows and their associated stresses have been an important
feature of dynamical models of the tachocline (e.g. Spiegel \& Zahn 1992,
Gough \& McIntyre 1998),
and such flows are generally associated with $P_2(\cos\theta)$ 
stresses (Schwarzschild 1958), what is striking here is the manifest 
quadrupolar morphology.
In \S 2.6 below we use explicit quadrupolar forcing to leverage a mathematically precise
isorotational contour solution from our fundamental equation, with 
essentially no other input.
So it is of some interest to try to understand its origin.  

The most important omission from the thermal wind equation, the
basis of our analysis, is the contribution from turbulent stresses and meridional
flow.  
These may be kinetic or magnetic, but they will generally take the form 
of a sum of terms proportional to  
$$
\dd_i(V_i W_\phi/R)
$$
where the index $i$ is itself summed over, and $V$ and $W$ represent symbolic
vector quantities (velocity fluctuations, vorticity, current density, etc.).
It is the $r$ and $\theta$ components of $i$ that are important
(the $\phi$ components vanish either explicitly or upon azimuthal averaging);
hence the earlier designation of the resulting stress as ``meridional.'' 
The net contribution of these correlated products evidently behaves 
as a large scale quadrupolar forcing.  Given the symmetry of our flow, it is not
surprising that the leading order departure from spherical structure 
is quadrupolar, but it is rather surprising how well this approximation works.              

\subsection {$F(\theta) = P_2(\cos\theta)$: an explicit solution}

We now combine the results of the previous section to obtain an explicit
mathematical solution for $\Omega$.  
Equation (\ref{tra2}) implies quite generally that
\beq\label{tra3}
\sin^2\theta_0 = {r^2\over r_0^2} \sin^2\theta - {2f'GM_\odot\over r_0^2
\gamma}
\left( {1\over r} - {1\over r_0}\right).
\eeq
Now at $r=r_T$ equation (\ref{tra2}) gives,
\beq\label{tra4}
\sin^2\theta(r_T) = 
{r_0^2\over r_T^2} \sin^2\theta_0 + {2f'GM_\odot\over r_T^2
\gamma}
\left( {1\over r_T} - {1\over r_0}\right).
\eeq
Then, substituting equation (\ref{tra3}) in (\ref{tra4}) leads to 
\beq
\sin^2\theta(r_T) = 
{r^2\over r_T^2} \sin^2\theta + {2f'GM_\odot\over r_T^2
\gamma}
\left( {1\over r_T} - {1\over r}\right),
\eeq
or
\beq\label{tra6}
\cos^2\theta(r_T) =
\left( 1  - {r^2\over r_T^2}\right) \sin^2\theta + {2f'GM_\odot\over r_T^2
\gamma}
\left( {1\over r} - {1\over r_T}\right),
\eeq
Then, equation (\ref{omg}) becomes 
\beq\label{frml}
\Omega^2(r, \theta) = \Omega_0^2(\sin^2\theta_0) +\Theta(r_T-r)(\Delta \Omega^2)
P_2[\cos\theta(r_T)]
\eeq
where $\sin^2\theta_0$ is given by equation (\ref{tra3}), $\cos\theta(r_T)$
by equation (\ref{tra6}), and we have assumed that $F$ is the
Legendre polynomial of order 2.  

Equation (\ref{frml}) is the desired explicit solution for $\Omega^2$,
but there is another way to formulate the content of our
simple problem:  given $\Delta\Omega^2$, what kind of a spread of surface
velocities is associated with this jump?  
The answer is
\beq
\Omega_0^2(\sin^2\theta_0) = \Omega_{rad}^2 - (\Delta \Omega^2) P_2[\cos\theta(r_T)]
= \Omega^2 - (\Delta \Omega^2) +{3\over2} (\Delta\Omega^2)\sin^2[\theta(r_T)]
\eeq
where $\Omega_{rad}$ is the angular velocity of the uniformly rotating radiative
interior.  
With the aid of equation (\ref{tra4}), we may rewrite this in terms of     
$\sin^2\theta_0$.  Choosing $r_0^2/r_T^2 =2$, and denoting
the final term of equation (\ref{tra4}) as $\xi$ (a number of order, but
typically less than, unity), we find
\beq\label{sun1}
\Omega_0^2(\sin^2\theta_0)  
= \Omega_{rad}^2 -(1+\xi)  (\Delta \Omega^2) +{3} (\Delta\Omega^2)\sin^2(\theta_0)
\eeq
This calculation is a bit too simple to generate more complexity than a $\sin^2\theta_0$
departure from uniform surface rotation.  Nevertheless, for reasonable values of $\Delta\Omega^2$,
equation (\ref{sun1}) gives a very respectable rendering of a solar-like 
surface rotation profile.  More importantly, it establishes a dynamical
coupling between
the spread in $\Omega^2$ present near the surface and   
the jump in $\Omega^2$ at the location of tachocline. 
The connection
between these two regions is possible because of the existence of linking
trajectory characteristics.   

Equation (\ref{sun1}) should not be construed as a statement that the spread in surface
angular velocities is actually {\it caused} by forcing from the tachocline; indeed, 
the sense of causality
is generally thought to run in the opposite sense: from the SCZ to the tachocline 
(e.g. Gough 2010).  Instead, equation (\ref{sun1}) shows explicitly
how the dynamics of thermal wind
balance plus external driving leads to a relatively simply coupling between the radial
jump in $\Omega^2$ at the tachocline and the angular spread in $\Omega^2$ at the surface.

\subsection {$f'(\Omega^2)$ linear in $\Omega^2$}

\subsubsection {Governing equation}

We next consider the case in which $f'$ varies linearly with
$\Omega^2$.  This is useful both in establishing that there is no particular
sensitivity to the assumption of a constant $f'$, and in allowing more detailed
structure to be revealed.  In fact,
the essential qualitative effects of a nonconstant $f'$ will be evident once
the governing equation is at hand.  

We begin by rewriting 
the trajectory characteristic equation (\ref{tra1}) in the 
compact form
\beq\label{ff2}
{dR^2\over dr} =  - {2g\over \gamma} f'(\Omega^2),
\eeq
where, as before, $R=r\sin\theta$, the axial radius.  
Let
\beq\label{f1}
f'(\Omega^2) = - \alpha -\beta\Omega^2,
\eeq
where $\alpha$ and $\beta$ are constants.  
Dividing by $g$,
differentiating with respect to $r$, and using equations (\ref{f1}) 
and (\ref{sol1}) leads to
\beq
{d\ \over dr} {1\over g}{d R^2\over dr} = {2\beta\over \gamma} T(r,\theta)
\eeq
Recalling that $g=GM_\sol /r^2$, this leads to the final form of our equation
for the trajectory characteristics, now free of all explicit $\Omega$ dependence:
\beq\label{R21}
{d^2\over dr^2}(rR^2) = {2\beta GM_\sol \over \gamma r} \, T(r,\theta).
\eeq
We now make the very reasonable and important assumption that {\em $T(r, \theta)$ is
local in $r$,} so that this variable is always near the shell $r=r_T$ whenever $T$ is
finite.  In essence, we ignore curvature terms, something we must do self-consistently
since these global terms are not known for general $T(r,\theta)$.  The local
model avoids this difficulty while retaining the essential physics, because the
tachocline transition is not, in reality, very broad.  

Consider equation (\ref{R21}) when the driving term $T(r,\theta)$ takes
the (radially local) form
\beq
T(r,\theta) = - T_0 P_2(\cos\theta)
\eeq
where $T_0$ is a constant term.  We assume that $T$ is present only for
$r<r_T$.  In this region, the differential equation (\ref{R21}) becomes
\beq\label{ode3}
{d^2(r^3\sin^2\theta)\over dr^2}  = {2\beta GM_\sol T_0\over \gamma r}
\left( {3\over 2} \sin^2\theta -1\right)
\eeq
It should be noted that the function $f'$ makes its appearance here only through
the $\beta$ parameter, and then only via the combination $\beta T_0$.  This
illustrates the point first made in the Introduction: 
a different value of $\beta$ can equivalently be regarded as keeping
$\beta$ the same and changing $T_0$.   There is nothing to be gained by allowing
a change in $f'$ in the tachocline, at least not for the class of model of
interest here.  

Replacing $r$ by $r_T$ everywhere leads to the differential equation
\beq
{d^2(\sin^2\theta)\over dr^2}  = {3\beta GM_\sol T_0\over \gamma r^4_T}
\left( \sin^2\theta -{2\over3}\right)\equiv k^2
\left( \sin^2\theta -{2\over3}\right)
\eeq
This is a {\em linear} second order equation for 
$\sin^2\theta$, and the solution is
\beq\label{aa7}
\sin^2\theta = A_1\cosh(kr) + A_2 \sinh (kr) +{2\over 3}
\eeq
where $A_1$ and $A_2$ are integration constants to be determined\footnote{
Equation (\ref{ode3}) may in fact be solved analytically without making
the local approximation.  But since locality has  been used implicitly
in ignoring the radial structure in $T$, it is best to be self-consistent.}.

\subsubsection {Global solution}

The characteristic solution for $r>r_T$ is already known (BBLW):
\beq\label{aa6}
\sin^2\theta = {r_0^2\over r^2} \sin^2\theta_0 + {2GM_\sol f'(\Omega_0^2)\over\gamma r^2}
\left( {1\over r}-{1\over r_0}\right)
\eeq
where 
\beq\label{aaaa}
\Omega_0^2(\sin^2\theta_0)= \Omega_1^2 +\Omega_2^2\sin^2\theta_0,
\eeq
and $\Omega_1^2$ and $\Omega_2^2$ are fixed constants.

We seek global solutions in which the exterior solution (\ref{aa6}) for $r>r_T$
joins smoothly to the interior solution (\ref{aa7}) for $r<r_T$ at $r=r_T$.
The solution for $\sin^2\theta$ and its first derivative with respect to $r$ should
be continuous.  Two dimensionless parameters are needed, defined by
\beq
\xi_1 = {2GM_\sol \beta \Omega_2^2\over \gamma r_T^3},
\eeq
\beq
\xi_2 = {2GM_\sol (\alpha +\beta\Omega_1^2)\over \gamma r_T^3}.
\eeq
An easy way to determine representative values of $\xi_1$ and $\xi_2$
is to use the $B$ parameter introduced in B09 for solving for the form of
the SCZ isorotation curves:
\beq\label{out1}
- {B\over r_0^3}\equiv - {2GM_\sol f'(\Omega_0^2) \Omega_2^2\over \gamma r_0^3}
\eeq
(Note that the solar radius notation $r_\sol$ was used in B09 instead of our $r_0$.)
A very simple and reasonable SCZ model is discussed in B09 using
\beq
- {B\over r_0^3} = 0.12 + 0.8\sin^2\theta_0,
\eeq
which would imply
\beq
\xi_1 = \left(r_0\over r_T\right)^3 0.8, \qquad 
\xi_2 = \left(r_0\over r_T\right)^3 0.12.
\eeq
Two other parameter combinations may be defined that will be later useful:
\beq
\xi_3 = \left(r_0\over r_T\right)^2 -\left[ 1 - \left(r_T\over r_0\right)\right] \xi_1
\eeq
\beq
\xi_4 = \xi_2 \left[ 1 - \left(r_T\over r_0\right)\right]
\eeq

At $r=r_T$, the second derivative of $\sin^2\theta$ along a characteristic is discontinuous,
but its value and first derivative are continuous.  With equations 
(\ref{aa7}) and (\ref{aa6}), together with the $\xi$ parameters, these two continuity
conditions may be written:
\beq
\xi_3\sin^2\theta_0 -\xi_4 = A_1\cosh kr_T + A_2\sinh kr_T +{2\over3},
\eeq
\beq
(\xi_1-2\xi_3)\sin^2\theta_0 +\xi_2+2\xi_4 = kr_T(A_1\sinh kr_T + A_2\cosh kr_T)
\eeq
Solving for $A_1$ and $A_2$, we find the following
interior solution for the characteristics:
\beq\label{csol}
\sin^2\theta -{2\over 3} = [\xi_3\sin^2\theta_0 -\xi_4 -{2/3}]\cosh k(r_T-r)
-{\sinh k(r_T-r)\over kr_T} 
\left[ (\xi_1-2\xi_3)\sin^2\theta_0 +\xi_2 + 2\xi_4 \right]
\eeq
For future reference, we list here the form of this characteristic isolating
$\sin^2\theta_0$:
\beq\label{sinth0}
\sin^2\theta_0 = {
\sin^2\theta -2/3 + (\xi_4+2/3)\cosh k(r_T-r) +\sinh k(r_T-r)(\xi_2+2\xi_4)/kr_T
\over 
\xi_3\cosh k(r_T-r) +\sinh k(r_T-r)(2\xi_3-\xi_1)/kr_T}
\eeq

To solve for $\Omega^2$ along the characteristic (\ref{csol}), use equation
(\ref{sol2}), noting that $T(r, \theta)$ vanishes for $r_0>r>r_T$:
\beq
\Omega^2 = \Omega_0^2 - \int^{r_T}_r T(r,\theta) dr = 
\Omega_0^2 +{3T_0\over 2} \int^{r_T}_r (\sin^2\theta(r) - 2/3) dr
\eeq
Integrating with the help of equation (\ref{csol}) we obtain
$$
\Omega^2  = \Omega_0^2 + {3T_0\over 2}\left[
\big((\xi_1-2\xi_3)\sin^2\theta_0  +\xi_2+2\xi_4\big)
(1/k^2r_T)(\cosh k(r_T-r) -1)\right. \qquad\qquad
$$
\beq\left.
\qquad\qquad -(\xi_3\sin^2\theta_0 -\xi_4-2/3)(1/k)\sinh k(r_T-r)
\right]
\eeq
This gives us the solution along the characteristic beginning at the surface value
$\Omega_0^2$, which is itself a function of $\sin^2\theta_0$.   We would like to
have an expression for $\Omega^2(r,\theta)$, without reference to the characteristics,
so that we may obtain the isorotation contours.  To this end, we first use 
equation (\ref{aaaa})
to expand $\Omega^2_0$, followed by equation (\ref{sinth0}) to eliminate $\sin^2\theta_0$
from the above expression.  In this way, we may obtain the desired isorotation
contours in the form of $\sin^2\theta$ as a function
of $\Omega^2$ and $r$ (as well the various parameters of course).  The result of this 
rather lengthy algebraic excursion is that the isorotation contours in the 
tachocline are given by:
\beq\label{isorot}
\sin^2\theta - {2\over 3} = 
{H_1 - H_2 +H_3\over 1-H_4}
\eeq
where 
\beq
H_1 = {\Omega^2-\Omega_1^2\over \Omega_2^2}\left[ \xi_3\cosh k(r_T-r) +
(2\xi_3-\xi_1){\sinh k(r_T-r)\over kr_T} \right]
\eeq
\beq
H_2=  \left[
(\xi_4+2/3)\cosh k(r_T-r) + (\xi_2+2\xi_4)(1/kr_T)\sinh k(r_R-r)
\right]
\eeq
\beq
H_3 = \left[ \xi_3(\xi_2+ 2 \xi_4)-(\xi_2+2/3)(2\xi_3-\xi_1)
\right] {\cosh k(r_T-r) -1\over\xi_1 }
\eeq
\beq
H_4 = {kr_T\over\xi_1}\left( \xi_3\sinh k(r_R-r) +{(2\xi_3-\xi_1)(\cosh k(r_T-r)
-1)\over kr_T} \right)
\eeq
Equations (\ref{csol}) and (\ref{isorot}) for the characteristics and isorotation
contours respectively comprise the interior tachocline solution for $\Omega(r, \theta)$.  

Notice that we have not made use of any inner boundary condition enforcing uniform
rotation at a particular location marking the beginning of the radiative zone.  In principle,
this location is part of what determines the form of the $T(r,\theta)$ stress.   
In practice,
however, the simple model that we have adopted reproduces the helioseismology data so
well that we may we just stop the calculation at a value of $r$ near $0.7$.  The isorotation
contours at this point are very nearly spherical (cf. figure [4]).  

\subsubsection {Modeling the Sun}

At a transition radius of $r_T=0.77 r_0$,  
the helioseismology data show a modification of what we refer to
in this work as the ``exterior'' solution.   
This translates to 
\beq
\xi_1 = 1.75233, \quad \xi_2=0.26285, \quad \xi_3=1.28539, \quad \xi_4=0.060456
\eeq
The quantity $(\Omega^2-\Omega_1^2)/\Omega_2^2$ is simply $\sin^2\theta_0$, where
$\theta_0$ is the surface colatitude angle of the isorotation curve.  Plots
of the resulting characteristics and isorotation contours are shown in figures
(3) and (4) for the case $kr_T=3$.  

\begin{figure}
\epsfig{file=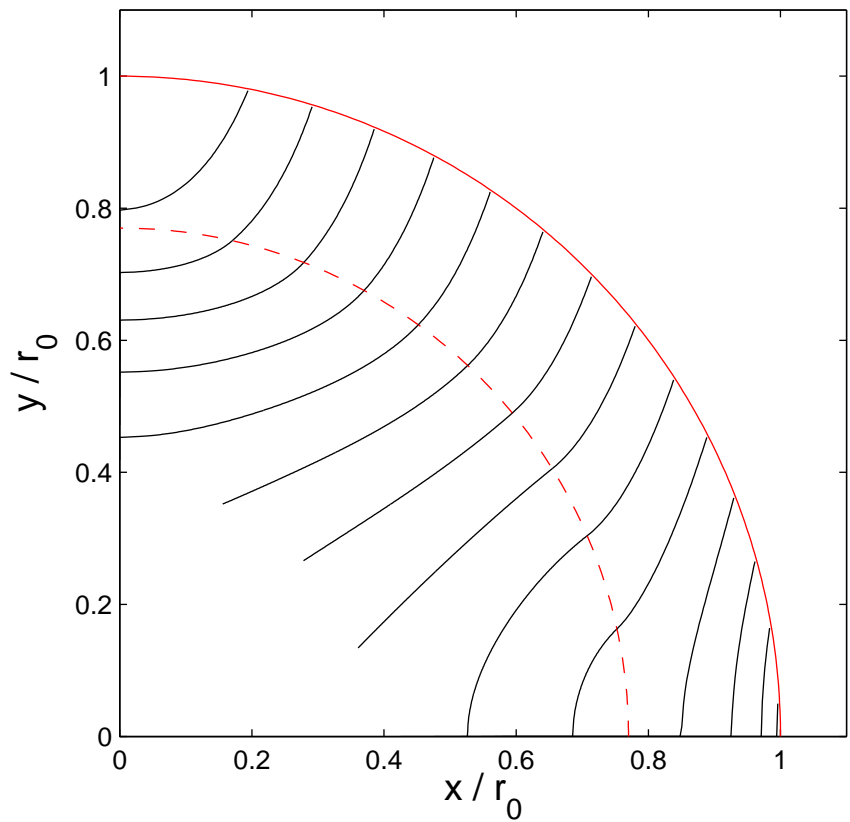}
\caption{Trajectory characteristics (eq. [\ref{csol}]) 
for the SCZ and tachocline.  The transition
radius, shown as a dotted line, is 0.77 of the surface radius,
and we have chosen $kr_T=3$.  
The curves, which are here displayed well beyond their relevant physical
domain, are
everywhere regular and have continuous first derivatives at $r=r_T$.}  
\end{figure}
\begin{figure}
\epsfig{file=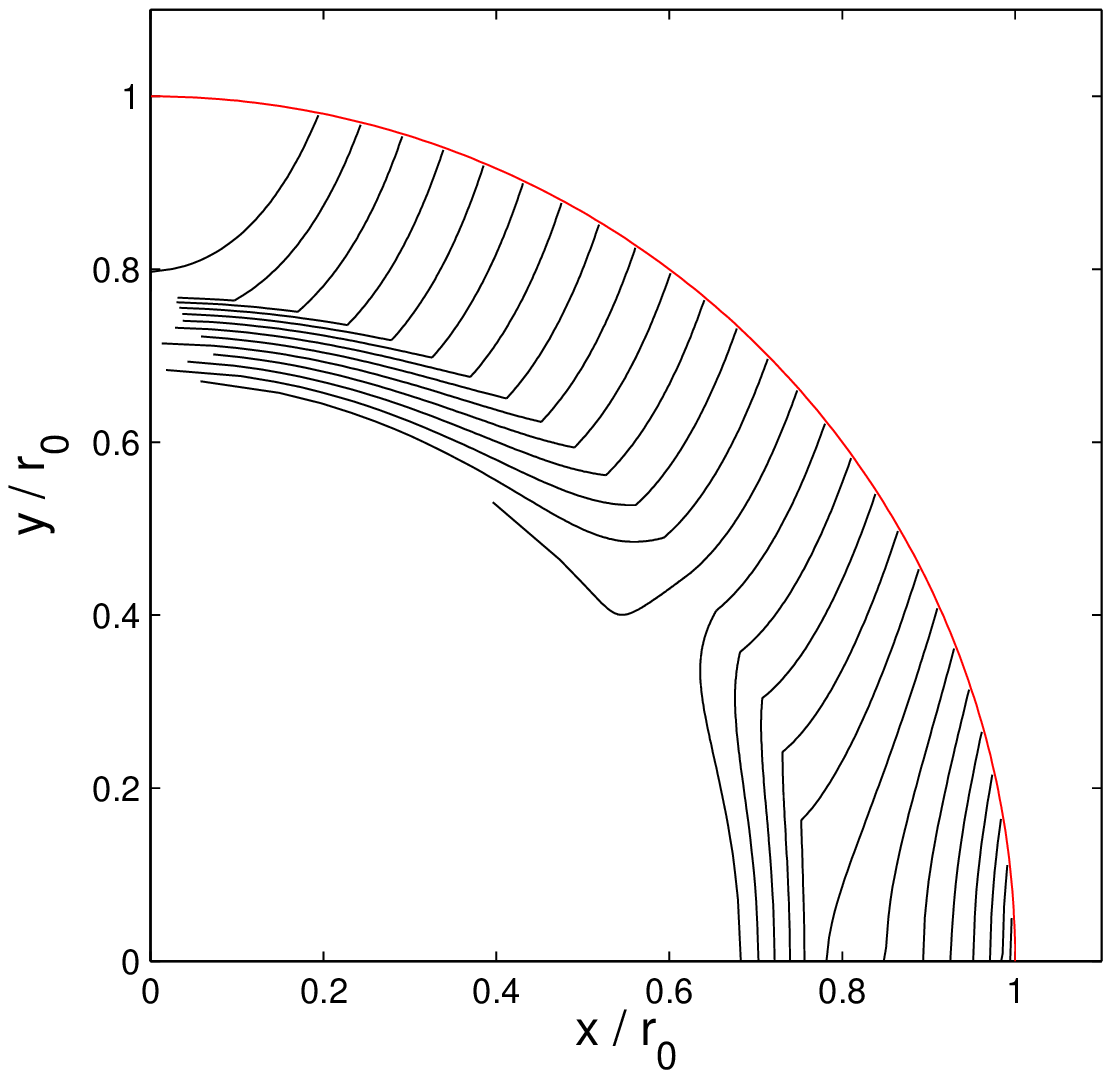}
\caption {Isorotation contours (eq. [\ref{isorot}]), with
$r_T=0.77 r_0$, $kr_T=3$.  The curves are continuous at $r_T$ but have discontinuous
first derivatives.  The ensemble of isorotational contours bears a clear and 
striking resemblance to the helioseismology GONG data (cf. figure [2]).} 
\end{figure}
                
These figures are remarkable.  The characteristics in figure (3) are everywhere smooth
and regular.  They are identical to isorotation contours for $r>r_T$, and bear no
resemblance to them for $r<r_T$.  The corresponding isorotation contours
(figure [4]) are continuous,
but have sharply discontinuous first derivatives at $r=r_T$.  What is astonishing,
however, is the extent to which they resemble the helioseismology GONG data
(cf. figure [2]).
Recall that our model consists, in its entirety, of i) the
assumption of cospatial surfaces of constant residual entropy
and angular velocity in the thermal wind equation; and ii) 
local spatial forcing via an inhomogeneous term 
proportional to $P_2(\cos\theta)$ that turns on at $r=r_T$.  
There is nothing else.   Indeed, to keep the mathematics as simple as
possible, we have not even used the best fit for the exterior
solution shown in figure (2), but rather something much more
basic.  Still, the similarity of our closed form solution to
the GONG data is unmistakable. 

Our results can be compactly summarized: the analysis shows that the
SCZ is essentially in thermal wind balance from near the surface down to a transition
radius of $r_T=0.77 r_0$, at which point distinctly quadrupolar tacholine meridional
stresses bring the rotation profile to that of a solid body.  
For all of its detailed complexity, the role of turbulent angular momentum 
transport in the SCZ is strictly
constrained, much as thermal transport
has little choice but to establish a nearly adiabatic temperature profile.  

\section {Conclusion}
The level of dynamical detail that can now be
elucidated in numerical simulations of the SCZ is
truly impressive (e.g. Miesch et al. 2008),  
but simulations incorporating
the tachocline into global models of solar differential 
rotation are only just beginning to take form 
(Browning et al. 2006, Miesch et al. 2009).  
The solar tachocline remains a complex, poorly understood 
region of turbulent activity.   While the details of the
interaction between the tachocline and the large scale differential rotation of
the SCZ are not yet within reach, certain
gross features of this interaction
may in fact be tractable.  In this work,             
we have made an inroad by assuming that entropy
is well-mixed in surfaces of constant angular velocity
within the convectively unstable portion of the tachocline, 
just as it appears to be within the SCZ.  Since the convective
Rossby number is likely to be even smaller here than in the SCZ, 
this would seem to be a reasonable approximation.  

The explicit separation of the more general tachocline stresses
from the mean inertial and baroclinic terms  
in the governing equation of vorticity 
conservation in its developed thermal wind form, and their
reduction to a single formal inhomogeneous forcing term, has proven to be
mathematically advantageous.
The precise nature of this stress is, of course, of an
unknown character.  We would
argue, however, that its consequences are not {\em unknowable.}
The stress operates locally
in radius and exhibits a definite quadrupolar angular structure: this is a very     
important point we wish to emphasize. 
It can be put to good use, because the  
formalism of quasilinear partial differential equations
may be brought to bear on the problem.   In particular, one
may analyze the problem in terms of trajectory characteristics,
whose form depends only implicitly on the tachocline stress terms, and
solution characteristics, along which $\Omega$ 
changes explicitly due to these same terms.  

In BBLW theory, the solution characteristics convey 
constant $\Omega$ along the trajectory characteristics.  
These characteristics are then identical to the isorotation contours.  
It is thus of considerable interest that in a simple and revealing       
class of solutions, the trajectory characteristics are mathematically
identical to those of BBLW theory, even within the tachocline.  
The solution characteristics are certainly altered, however: they
dictate that $\Omega$ is initially constant along the
trajectories, but then changes rapidly along these same curves as
the tachocline is penetrated from above.  
The result is quasi-spherical isorotation contours inside of the tachocline,
followed by a sharp upturn when the convection zone is reached from below.  
This behavior is also clearly seen in a more sophisticated, but still explicit,
solution that is presented in this paper (cf. figure [4]).  

The radially local nature of the tachocline stresses
is the crux of the problem, as it allows
for mathematical modeling independent of the need of detailed
knowledge of the spatial structure of the forcing.  The simplest approach
is to use Green function techniques with forcing
by a Dirac $\delta$-function at a transition radius $r_T$.   
The dominant $P_2(\cos\theta)$
angular structure of the tachocline stresses becomes very
apparent when the radial behavior is simplified in this way.
Since the monopole response is evidently
small, a reasonable surmise is that the turbulent stresses in the
tachocline are driven by off-diagonal terms in a correlation tensor.  

Angular momentum stresses within the SCZ create differential rotation,
even if the initial condition is one of uniform rotation.   
Thus, in a causal sense, the tachocline likely owes
its existence to the stresses of the convective zone, and not vice-versa.
In its current state of development, our calculation cannot directly address this 
causal link, but it does suggest how a dynamical couple arises 
between the surface spread in angular
velocity and the tachocline angular velocity jump.  
This characteristic-based connection
between the jump in the angular velocity at the tachocline and the spread
of angular velocities at the solar surface is an important consequence of
simple quadrupolar forcing.   

%

Venturing even a little beyond a simple Green's function approach gives a
much richer return.  If one allows both an $\Omega$-dependence in our $f'$ function
and a finite thickness for the transition region,  
the mathematical structure of the problem becomes more complicated:
the solution and trajectory characteristics are now coupled.   But it is still possible
to obtain an explicit closed form solution for the isorotation contours.  This,
the principal result of this paper, is shown in figure (4).  The relatively
simple solution so much resembles the GONG contours
seen in figure (2), there can be little doubt that at the
very least, the mathematical content
of our approach is sound.  Even if viewed at a purely phenomenological
level, this is surely progress.  

While our results provide a simple framework 
to aid in the elucidation of the global character of solar differential rotation,
many questions remain unanswered.  Quite apart from the grand
challenge problem of tachocline structure and
dynamics, it is legitimate to question the
work presented here on its own terms.  How general and robust, for example,
is the separation embodied in equation (5), or the critical assumption 
that a functional
relation remains between residual entropy and angular momentum in the
lower buoyancy tachocline?  Why should such a simple approach work so
well?   There is also evidence that the tachocline thickness 
varies with $\theta$ (Christensen-Dalsgaard \& Thompson 2007),
an effect completely neglected here.  The relative simplicity of our 
arguments is somewhat deceptive, because it masks the deeper
conceptual problem of understanding precisely {\em how}
the coupling between the tachocline and the convective zone rotation
profile is established (see, e.g. Rempel 2005, Gough 2010).  
These issues certainly have not been addressed by this
work; doing so without recourse to turbulence modeling will remain the
province of numerical simulation for the foreseeable future. 

But in this weakness also lies some power.  We have 
noted earlier that the SCZ adiabatic temperature profile sheds little light
on the dynamics of convective transport, though it is in an important
sense completely beholden to it.  This is not usually
viewed as a shortcoming; rather it is a useful and powerful
constraint.  This is a point worth remembering: some aspects of 
the solar rotation problem are bound to be
less mysterious than others.  If thermal wind balance
and the confluence of isorotation and constant residual entropy surfaces
lead to important constraints for the Sun's rotation profile, the 
theorist's task will be greatly simplified---and understanding how 
the Sun rotates may prove to be a less daunting challenge than it might 
have first appeared.

\section*{Acknowldgements}
This work benefited from an exceptionally thorough and constructive
review by our referee
Mark Miesch, and we are most grateful to him for many valuable comments.
We are likewise indebted to Nigel Weiss for important advice in the early stages
of this work.  Finally, we thank Rachel Howe for
making reduced GONG data available to us in electronic form.  This work has
been supported by a grant from the Conseil R\'egional de l'Ile de France.

\end{document}